\documentclass[aps,pra,twocolumn]{revtex4}
\usepackage{graphicx}
\usepackage{bm}
\usepackage{color}
\usepackage{wasysym}

\usepackage[applemac]{inputenc}

\begin{document}
\preprint{}
\title{Observed quantum dynamics: classical dynamics and lack of Zeno effect}
\author{Juli\'an L\'opez, Laura Ares and Alfredo Luis}
\email{alluis@fis.ucm.es}
\homepage{http://www.ucm.es/info/gioq}
\affiliation{Departamento de \'{O}ptica, Facultad de Ciencias
F\'{\i}sicas, Universidad Complutense, 28040 Madrid, Spain}
\date{\today}

\begin{abstract}
We examine a case study where classical evolution emerges when observing a quantum evolution. By using a single-mode quantum Kerr evolution interrupted by measurement of the double-homodyne kind (projecting the evolved field state into classical-like coherent states or quantum squeezed states), we show that irrespective of whether the measurement is classical or quantum there is no quantum Zeno effect and the evolution turns out to be classical.
\end{abstract}


\maketitle

\section{Introduction}
The proper relation between the quantum and classical theories has been a subject of interest, research and debate from the very beginning of the quantum theory, specially the quantum measurement processes and their effects. We may say that at a fundamental level the issue is not solved yet. 

We may refer to some rather formal mathematical limits such as $\hbar \rightarrow \infty$ specially inspiring if applied to coherent and Gaussian states, and in general classical-like states \cite{HA80,BA90}. Regarding more practical approaches, the most popular account refers to decoherence as the practical mechanism by which quantum paradoxes disappear leading to the emergence of the classical world \cite{MH86,HSZ98,BHJ00,SM01}. 

Decoherence is the result of the coupling of the system with a large enough environment: both system and environment are modified by this coupling in different forms according to their different size and complexity. The key point for us is that this is the basis of measurement on its more pure form. Inspired by this reasoning, in this work we follow a promising avenue of research which may be formulated in this way: Classical dynamics are just observed quantum dynamics \cite{SM01}.

We prove this idea in a very specific arena. We use a nonlinear single-mode Kerr effect \cite{ST91}, which produces notable quantum phenomena, such as revivals and Schr\"{o}dinger cat states \cite{YS86,MT87,RK13}. The evolution is observed via a complex-amplitude measurement of the kind of double-homodyne detection that, depending on its balanced or unbalanced setting, is governed by projection on classical-like coherent states or quantum squeezed states, respectively \cite{NW87,WC84,WC86}. The idea is that these measurements may be fuzzy enough to respect the (classical) details of the evolution. In this regard, coherent and squeezed states form a variety isomorphic to the phase space, in our case a plane.

Frequent measurements checking the state of the system is the usual arena for the appearance of the Zeno effect. However, this does not occur in this model. This is analyzed in some detail in Sec. 4 examining which characteristic of the observation procedure is able to impede the appearance of quantum features while inhibiting the appearance of the Zeno effect.

\section{Unobserved classical and quantum dynamics}

\subsection{Unobserved classical dynamics}

The Hamiltonian describing the nonlinear part of a single-mode propagation through a Kerr medium is of the form \cite{ST91}:
\begin{equation}
\label{cH}
    H_c = \chi |\alpha|^4, 
\end{equation}
where $\chi$ is the corresponding nonlinear susceptibility in appropriate units, and $\alpha$ is the dimensionless complex amplitude of the field mode. For definiteness we consider the interaction picture where we focus just on the effects caused by the nonlinear term (\ref{cH}) that results in 
\begin{equation}
\label{ce}
\alpha(t) = \alpha e^{-i\Omega t}, \qquad \Omega = 2\chi |\alpha|^2 \;,
\end{equation}
where it must be noticed that $|\alpha|$ is a constant of the motion. 

\bigskip

\subsection{Unobserved quantum dynamics}

The quantum version of the Hamiltonian (\ref{cH}) is, in units $\hbar =1$,
\begin{equation}
H =  \chi \hat{n}^2 , \qquad \hat{n} = a^\dagger a ,
\end{equation}
where $\hat{n}$ is the photon-number operator, and $a$ is the complex-amplitude operator satisfying the commutation relation $[a,a^\dagger] = 1$. In the quantum case we find advantageous to express the evolution via the action of the unitary operator 
\begin{equation}
U(t) = e^{-it \chi \hat{n}^2} .
\end{equation}
For the sake of illustration we may consider the evolution of the mean value of $a$ when the field is initially in a Glauber coherent state $| \alpha \rangle$, defined by the eigenvalue equation $a |\alpha \rangle = \alpha |\alpha \rangle$ \cite{RG63a,RG63b,ES63,MW95,SZ97}. The result is 
\begin{equation}
\langle a \rangle  = \alpha e^{-2|\alpha|^2\sin^2(\chi t)}e^{-i[\chi t + |\alpha|^2 \sin(2\chi t)]} .
\end{equation}
In Fig. 1 we represent the evolution of the real part of $\langle a \rangle$ for $\alpha = 4$ and $\alpha = 1$, showing the structure of collapses and revivals as a consequence of the creation of Schr\"{o}dinger cat states \cite{YS86,MT87,RK13}.

\begin{figure}[h]
    \centering
    \includegraphics[width=8cm]{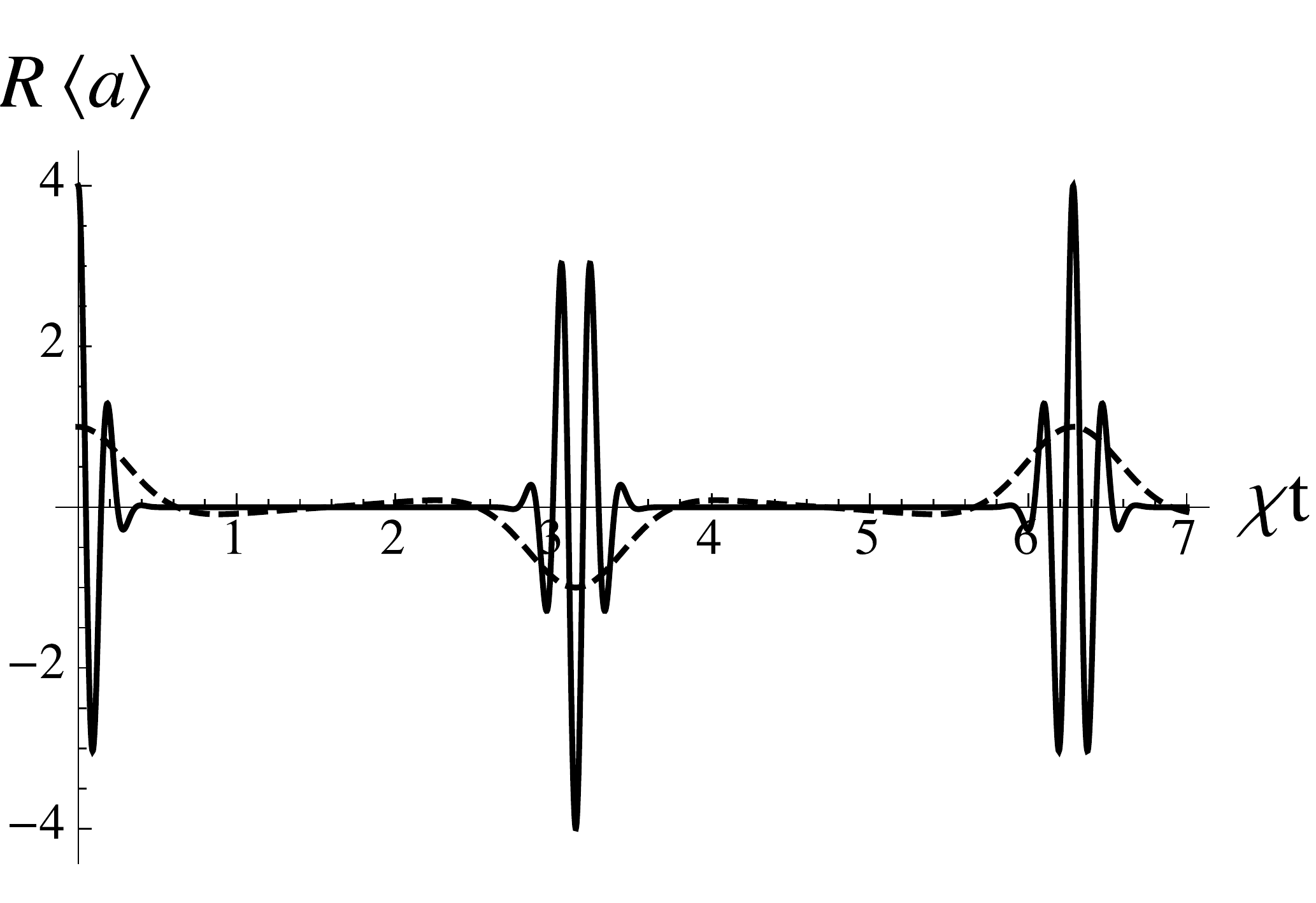}
    \caption{Collapses and revivals on an unobserved Kerr evolution of the real part of $\langle a \rangle$ as a function of $\chi t$ for an initial coherent state with $\alpha = 4$, solid line, and $\alpha = 1$, dashed line.}
\end{figure}

\section{Observed quantum evolution}

\subsection{Measurement}

In the following we may find convenient the decomposition of $a$ in terms of quadrature operators $\hat{q}$, $\hat{p}$, satisfying the typical position-linear momentum commutation relation
\begin{equation}
a = \frac{1}{\sqrt{2}}(\hat{q}+i \hat{p}) ,\quad [\hat{q}, \hat{p}] = i. 
\end{equation}

As a suitable measurement we will consider the projection on displaced states
\begin{equation}
\label{DStates}
    | z \rangle =  e^{\alpha a^\dagger-\alpha^\ast a}  | \psi \rangle , \quad z = \pmatrix{q \cr p} , \quad \alpha = \frac{1}{\sqrt{2}}(q+i p) 
\end{equation}
where $| \psi \rangle$ can be in principle any state. For the whole procedure it is crucial that the state-labels, either in the vector form $z$, or in the complex scalar $\alpha$, form a variety isomorphic to the phase space of the observed system. Thus, the outcomes $z$ can be regarded as an observation of the phase space.

If $| \psi \rangle = |0 \rangle$ is the vacuum state, then $| z \rangle$ are the Glauber coherent states universally considered as the most classical-like pure states \cite{RG63a,RG63b,ES63,MW95}. If $| \psi \rangle $ is a squeezed vacuum state, then $| z \rangle$ are squeezed coherent states, and clearly nonclassical \cite{SZ97}. So, this simple model includes projection on classical or nonclassical states. We recall that for  any $| \psi \rangle $ the states $| z \rangle$ provide a resolution of identity \cite{AC96}
\begin{equation}
\label{ri}
    \frac{1}{2\pi} \int d^2 z | z \rangle \langle z | = I,
\end{equation}
where $ d^2 z  = dq dp$, and $I$ is the identity.

\subsection{Process}

Without loss of generality, we consider that the initial sate belongs to this same measurement family $| z \rangle$, and will be denoted by $|z_0 \rangle$.  Otherwise, the evolved state will be forced to be one of the family $| z \rangle$ by the projection associated to the very first measurement. The quantum evolution from $t=0$ to $t$  is interrupted $N$ times  at times $t_j = j \tau$, $j=1,2,\ldots N$, to perform a measurement whose effect is the projection of the system state on some vector $| z_j \rangle$. Therefore we have a series of $N$ continuous evolution during a time $\tau$ governed by the action of the unitary operator interrupted by $N$ sudden jumps from $U(\tau) | z_j \rangle$ to  $| z_{j+1} \rangle$. 

The fundamental quantity regarding the observed evolution is the conditional probability
\begin{equation}
\label{bc}
p(z_{j+1} | z_j) = \frac{1}{2\pi} \left | \langle z_{j+1} | U(\tau) | z_j \rangle \right |^2 .
\end{equation}

The transition probability $p(z_{j+1} | z_j)$ from  $z_j$ to $z_{j+1}$ does not depend on the preceding results $z_{j-1}, z_{j-2}, \dots$ and consequently the process is Markovian as far as there is no memory. This Markovian character is enforced by the quantum reduction projecting the evolved state into one of the states $|z\rangle$ depending just on the outcome and erasing any previous information about the evolution.

In most situations, we will not be interested in keeping track of the intermediate results, being just interested in the final distribution for the last outcome $z=z_N$ after a total evolution time $t=N \tau$, which is 
\begin{equation}
\label{pzz0}
p_N (z | z_0 ) = \int d^2 z_1 \cdots d^2 z_{N-1}  p(z | z_{N-1}) \cdots p(z_1 | z_0) .
\end{equation}

\bigskip

Roughly speaking, we may imagine that the result is some random deviation from a mean drift caused by the Hamiltonian part of the evolution. The fundamental question to be addressed here is whether this drift resembles the classical evolution or not, and if so, which is the particular effect of the measurement. 

\bigskip

\subsection{Linear approximation}

Any progress along this line involves the computation of the transition probability $p(z_{j+1} | z_j) $ in Eq. (\ref{bc}). This is in general rather awkward unless some suitable approximations are allowed. We consider not too quantum and intense enough fields, so that $\bar{n} \gg 1$ and $\Delta n/\bar{n} \ll 1$ that allows us to approximate $\hat{n}^2$ by 
\begin{equation}
\label{la}
\hat{n}^2 \approx \bar{n}^2 + 2\bar{n}(\hat{n} - \bar{n}) + \dots \approx 2\bar{n}\hat{n} - \bar{n}^2 ,
\end{equation}
where $\bar{n}$ is the mean number of photons of the initial state, which is a constant of the motion, so  that when $\tau \rightarrow 0$ 
\begin{equation}
 U(\tau) \simeq e^{- 2i \chi \bar{n} \tau \hat{n} },
\end{equation} 
and the evolution becomes a linear transformation
\begin{eqnarray}
\label{qe}
&  U^\dagger (\tau) \hat{z} U(\tau) \simeq M (\tau ) \hat{z}, & \nonumber \\
 & & \nonumber \\
 & M (\tau )=  \left ( \begin{array}{cc} \cos (\Omega  \tau ) & \sin (\Omega \tau ) \\ - \sin (\Omega \tau ) & \cos (\Omega \tau ) \end{array} \right ), &
\end{eqnarray}
where 
\begin{equation}
\label{Om}
\Omega = 2 \chi \bar{n} ,  \qquad \hat{z} = \pmatrix{\hat{q} \cr \hat{p}}   ,
\end{equation}
which is fully equivalent to the classical evolution (\ref{ce}).

\bigskip

\subsection{Observed evolution via measurement}

\bigskip

Next we analyze the effect of the measurements. To this end we prove by induction the following theorem in Appendix A: 
\begin{equation}
\label{theo}
p_N (z | z_0) =  W_N \left [ M^{-1} (t  ) z- z_0 \right ] ,
\end{equation}
where $W_N (z)$ are functions determined in Appendix A. The key point of the theorem is that  the complete dependence on the initial $z_0$ and final $z$ phase-space-like coordinates  is exclusively  of the form $M^{-1} (t) z- z_0$, which is precisely the classical evolution (\ref{ce}). The difference with the classical case is the dependency of the functional form $W_N (z)$ on the number of measurements $N$.  
 
\bigskip

This result includes classical as well as nonclassical measurements since the states $| z \rangle$ can represent paradigmatic classical-like coherent states, as well as clearly nonclassical states such as arbitrary squeezed states, or even displaced number states, or many other sophisticated nonclassical states. 

\bigskip

The differences regarding particular choices of  $| \psi \rangle$  means a different structure for the fluctuations around the classical trajectory represented by $W_N (z)$. This is the factor that includes the nonclassical features of the measurement, as far as such fluctuations are of quantum origin.  

\bigskip

\subsection{Gaussian states}

In the case of squeezed coherent states we show in Appendix A that
\begin{equation}
W_N (z ) = \frac{1}{2\pi \sqrt{ \det C_N}}  e^{- z^T C_N^{-1} z/2} ,
\end{equation}
where $C_N$ is the corresponding covariance matrix  obtained via a recursive relation 
\begin{equation}
\label{CN}
C_N  = \sum_{j=0}^{N-1} M^{-j} C_1 \left ( M^{-j} \right )^T ,
\end{equation}
being

\begin{widetext}

\begin{equation}
\label{Gs}
C_1 = \left ( \begin{array}{cc} \cosh(2 r) + \cos^2 (\Omega \tau) \sinh (2 r) &  \frac{1}{2} \sin (2 \Omega \tau) \sinh (2 r) \\    \frac{1}{2} \sin (2 \Omega \tau) \sinh (2 r) & \cosh(2 r) - \cos^2 (\Omega \tau) \sinh (2 r) \end{array} \right ).
\end{equation}

\end{widetext}

\noindent where $r$ is the compression parameter that can take any value between $-\infty$ and $\infty$.

\bigskip

The properties of the noise added by the observation can be estimated via $\det C_N$ which is the Robertson--Schr\"{o}dinger form of phase-space  uncertainty relations, being $\det C_N = 1/4$ the ultimate quantum limit.

\bigskip

In the case of a classical-like measurement given by projection on Glauber coherent states, this is $r=0$ in Eq. (\ref{Gs}), we get that $C_1 = I$ so after Eq. (\ref{CN}) we readily obtain
\begin{equation}
\label{CNC}
C_N =  N I,
\end{equation}
This shows clearly that the only effect of the observation is to increase the quantum uncertainty around the classical trajectory in a way proportional to the number of measurements $N$.

\bigskip

In the particular case of measurements performed on displaced squeezed states, i. e., $r \neq 0$ in Eq. (\ref{Gs})  we have that  $\det C_N$ in Eq. (\ref{CN}) can be computed analytically in all cases, although it leads to expressions too long to be useful. Fortunately, in some meaningful limits suitable approximations can be derived. We consider two limits. On the one hand $\Omega \tau \ll 1$, which is consistent with the linear approximation (\ref{la}). On the other hand, we are interested in the limit $\Omega t \gg 1$ so that the Hamiltonian evolution has time to develop itself, since the quantum behavior, i.e., revivals, appear after a significant number of oscillations have been performed. Combining $\Omega \tau \ll 1$ and $\Omega t \gg 1$ implies $N\gg 1$. In this case the leading term of the exact (\ref{CN}) gives 
\begin{equation}
     \sqrt{\det C_N} = N \cosh (2r) .
\end{equation}
So that the uncertainty increases with regard to the classical like observation in a factor depending on the compression parameter, in agreement with the quantum origin of these fluctuations.

\bigskip

\section{Contextual Zeno effect}

The situation where a dynamics is frequently interrupted to detect whether the state remains in the initial state typically leads to the Zeno effect \cite{MS77,AP93,HW97}.  Since our initial state $| z_0 \rangle$ belongs to the measurement family $| z \rangle$, we are actually checking whether the evolved state continues in the initial state $| z_0 \rangle$. This is the usual scenario that leads to Zeno effect in the form of a complete stop of the evolution freezing the system in the initial state $| z_0 \rangle$. But this does not occur in our case, the evolution is no stopped, the uncertainity increases with $N$ and the Wigner function spreads over a growing area of phase space. We think this may deserve a brief analysis. 

\bigskip

The key point is the nature of the family of states where the measurement projects. For example, let us consider a dichotomic measurement with just two projectors   
\begin{equation}
\label{dis}
\Delta (0) = | z_0 \rangle \langle z_0 | ,  \quad
\Delta (\neg 0)=I - | z_0 \rangle \langle z_0 | ,
\end{equation}
where  $| z_0 \rangle$ is the initial state. The standard Zeno analysis leads to a survival probability $P_0$ that tends to one as measurement tends to be more frequent, i. e.,
\begin{equation}
P_0 \gtrsim e^{ - \Delta^2 \hat{n}^2 (\chi t)^2/ N } ,
\end{equation}
so $P_0 \rightarrow 1$ as $N \rightarrow  \infty$, where the right-hand side is computed assuming that all measurement results confirm that the system is in state $| z_0 \rangle$.  
\bigskip

The situation is completely different if the projection on the initial state $| z_0 \rangle$  is embedded on a continuous family of nonorthogonal projectors, such as in the measurement we are considering in this work, this is 
\begin{equation}
\Delta (z) = \frac{1}{2\pi} | z \rangle \langle z | ,
\end{equation}
and naturally a key point is the factor  $1/(2\pi)$. Let us compute in this case the survival probability in the best possible scenario $\Omega t = 2 m \pi$ for integer $m$ so that $M^{-N} z_0- z_0=0$. After Eqs. (\ref{theo}), (\ref{WNG}) and (\ref{CNC}) for a measurement projecting on coherent states we get
\begin{equation}
P_0 = p_N (z = z_0 | z_0 ) = \frac{1}{2\pi \sqrt{ \det C_N}} \propto \frac{1}{N} ,
\end{equation}
so that $P_0 \rightarrow 0$ as $N \rightarrow  \infty$. 

\bigskip

During an infinitesimal evolution the state moves  from $| z_0 \rangle$ to an infinitesimally close state $U(\tau) | z_0 \rangle$, and it turns out that within the family $| z \rangle$ there is  a neighbour state $| z \rangle$  different from $| z_0 \rangle$ closer to the  infinitesimally evolved state $U(\tau) | z_0 \rangle$ than $| z_0 \rangle$. This never happens if the measurement states are not so close enough, say as in Eq. (\ref{dis}). This is to say that the Zeno effect is very sensitive to the way in which the projection on the original state $| z_0 \rangle \langle z_0 |$ is embedded. In this way we may say that the Zeno effect is contextual.

\bigskip

\subsection{Overlapping kills the Zeno effect}
Let us try to elucidate further the reasons explaining the lack of Zeno effect. Regarding the measurement basis  $| z \rangle$ there are multiple characteristics that might contribute, such as continuity, overcompleteness, and overlapping between different  $| z \rangle$. We can present an extremely simple scenario with discrete outcomes and without overcompleteness, where the only explanation for the lack of Zeno effect is the overlap between the elements of the POVM.

\bigskip

Let us consider a two-dimensional space spanned by two orthogonal states $| 1 \rangle$, $| 2 \rangle$ and the following POVM, in such basis
\begin{equation}
    \Delta_1 =  \pmatrix{\cos^2 \alpha & 0 \cr 0 & 0}, \quad
 \Delta_2 = \pmatrix{\sin^2 \alpha & 0 \cr 0 & 1} ,
\end{equation}
assuming that the outcome-dependent reduced states after the measurement are the corresponding normalized states 
\begin{equation}
    \rho_1 = \pmatrix{1 & 0 \cr 0 & 0}, \quad  \rho_2 = \frac{1}{1+\sin^2 \alpha} \pmatrix{\sin^2 \alpha & 0 \cr 0 & 1}  ,
\end{equation}
being $\rho_1$ always the initial state. Note that the POVM is discrete, with just two only outcomes, so there is discreteness and no overcompleteness, while there is a clear overlap depending on the free parameter $\alpha$: 
\begin{equation}
    \mathrm{tr} \left (  \Delta_1 \Delta_2 \right ) = \frac{1}{4} \sin^2 ( 2 \alpha ) .
\end{equation}

\bigskip

Let the evolution be governed by the Hamiltonian, in units $\hbar = 1$,
\begin{equation}
    H =  \omega \pmatrix{0 & 1 \cr 1 & 0},
\end{equation}
so that 
\begin{equation}
    U(\tau ) = e^{-i H \tau} = \pmatrix{\cos (\omega \tau )  & - i \sin (\omega \tau ) \cr - i \sin (\omega \tau ) & \cos (\omega \tau ) } .
\end{equation}

\bigskip

The fundamental conditional probability reads in this case 
\begin{equation}
p(j| k) = \mathrm{tr}  \left [ \Delta_j U(\tau) \rho_k U^\dagger (\tau ) \right ] ,
\end{equation}
which can be suitably arranged in a $2 \times 2$ matrix as
\begin{equation}
T = \pmatrix{p(1|1) & p(1|2) \cr p(2|1) & p(2|2)},
\end{equation}
with
\begin{eqnarray}
& p(1|1) =  \cos^2 (\alpha ) \cos^2 (\omega \tau ), & \nonumber \\
& p(2|1) =  \sin^2 (\alpha ) \cos^2 (\omega \tau )+\sin^2 (\omega \tau ), & \nonumber \\
& p(1|2) =  \frac{\cos^2 (\alpha )\left [ \cos^2 (\omega \tau ) \sin^2 (\alpha )+\sin^2 (\omega \tau ) \right ]}{1+\sin^2 (\alpha )} , & \nonumber \\
& p(2|2) =  \frac{\cos^2 (\omega \tau )\left [ 1+ \sin^4 (\alpha ) \right ] +2 \sin^2 (\alpha )\sin^2 (\omega \tau )}{1+\sin^2 (\alpha )} .& 
\end{eqnarray}

\bigskip

The survival probability is  
\begin{equation}
\label{osp}
p_0 = \sum_{j,k, \dots, m = 1,2} p(1 | j)p(j |k) \cdots p(m | 1) ,
\end{equation}
which is actually the matrix element $(T^N)_{1,1}$. After some little algebra, the survival probability (\ref{osp}) becomes
\begin{equation}
\label{p0O}
    p_0 = \frac{\cos^2\alpha}{2} + \left( 1 - \frac{\cos^2\alpha}{2} \right) \left[ \frac{\cos^2\alpha\,\cos(2\omega\tau)}{2-\cos^2\alpha} \right]^N.
\end{equation}
When there is overlap, i. e., $\sin ( 2 \alpha ) \neq 0$, we get lack of Zeno effect, as far as for $N \rightarrow \infty$ only the first term survives:
\begin{equation}
    p_0\rightarrow \frac{\cos^2\alpha}{2},
\end{equation}
pointing to the overlap between POVM elements as the key feature inhibiting Zeno effect, the larger the overlap the lesser the survival probability. We also note that there is no classical-like evolution because the observation provides no enough density of states.  

\bigskip

To further investigate the interplay between Zeno and lack of Zeno effect, we may consider the situation in which the overlap may depend on the number of measurements $N$ through $\alpha$ in such a way that $\alpha \rightarrow 0$ as $N \rightarrow \infty$. When $N\gg 1$ and $\alpha \ll 1 $ we get that (\ref{p0O}) can be approximated as 
\begin{equation}
    p_0 \simeq \frac{1}{2} \left( 1 + e^{-2N \alpha^2} e^{-2\omega^2 t^2/N} \right ).
\end{equation}
So if $\alpha$ tends to $0$ faster that $1/\sqrt{N}$ Zeno effect occurs, while otherwise there is no Zeno effect.

\section{Conclusions}

We have presented a simple model where the observation of quantum dynamics leads to a fully classical evolution. We think there are some interesting points worth to be followed. We have obtained the same classical trajectory for classical as well as for quantum measurements. Nevertheless in general there are differences between both classes of observation in the structure and the amount of the uncertainty around the classical trajectory. This is interesting as far as such noise is purely quantum and introduced by the observation.  

There may be two basic features for the emergence of the classical dynamics presented in this work that might be further pursued in future research: i) Whether it is crucial that the manifold of measurement outcomes must be isomorphic to the phase space of the problem. ii) Whether it is crucial that the unobserved evolution can be well approximated linear transformations in the short time limit. Both points can be strongly dependent on the basic variables and operators used to parametrize both the measurement and the phase space. So we cannot exclude that these results might be universal under a suitable choice of variables adapted to the problem at hand.

Finally, we have analyzed the lack of Zeno effect in our model. In this sense, we have conclude that the cause of this absence is the overlap between the states where the measure projects. We support this conclusion with an example in which the overlap between POVMs elements is the only possible  cause. This dependence on the way of measuring is what makes us referring to the Zeno effect as contextual.

\bigskip

\section*{Acknowledgments}
L. A. and A. L. acknowledge financial support from Spanish Ministerio de Econom\'ia y Competitividad Project No. FIS2016-75199-P.
L. A. acknowledges financial support from European Social Fund and the Spanish Ministerio de Ciencia Innovaci\'{o}n y Universidades, Contract Grant No. BES-2017-081942.

\appendix

\section{Proof of theorem (\ref{theo})}

The basic transition probability (\ref{bc}) can be computed using the Wigner-function representation \cite{SZ97,HOSW84}. This is because of two key properties of Wigner functions:  i) Under linear transformations Wigner functions transform as classical probability distributions do by a simple transformation of arguments, say that for the transformation (\ref{qe}) we have  
\begin{equation}
\label{ct}
W(z;t) = W \left [ M^{-1} (\tau) z;0 \right ] ,
\end{equation}
so the Wigner functions of the vectors $|z_{j+1}\rangle$ and  $U(\tau) | z_j \rangle $ in Eq. (\ref{bc}) are, respectively, 
\begin{equation}
W(z-z_{j+1}), \qquad W\left [ M^{-1} (\tau ) z-z_j \right ] ,
\end{equation}
where $W(z )$ is the Wigner function of the state $|\psi \rangle$. ii) The scalar product of vectors can be computed by the overlap of their Wigner functions, i. e.,
\begin{equation}
\left | \langle \varphi | \psi \rangle \right |^2 = 2 \pi \int dz W_\varphi (z) W_\psi (z).
\end{equation}

Therefore the transition probability (\ref{bc}) can be expressed as 
\begin{equation}
p(z_{j+1} | z_j) =  \int d z W(z-z_{j+1} ) W \left [ M^{-1} (\tau )z- z_j \right ] ,
\end{equation}
so we are ready to compute  $p(z | z_0) $ via the chain in Eq. (\ref{pzz0}). 

\bigskip

With this we can prove theorem (\ref{theo}) by induction
\begin{equation}
p_N (z | z_0) = W_N \left [ M^{-N} (\tau ) z- z_0 \right ] ,
\end{equation}
where $W_N (z)$ are functions to be determined. 

\bigskip

So we begin with the first link in the chain (\ref{pzz0}), this is (\ref{bc}), 
\begin{equation}
p(z_1 | z_0) = \int d z W(z -z_1 ) W (M^{-1}z- z_0 )  ,
\end{equation}
and for simplicity we skip the dependence of $M$ on $\tau$. We perform the unit-Jacobian change of variables $z^\prime = z -z_1$ to get 
\begin{equation}
p(z_1 | z_0) = \int d z^\prime W(z^\prime ) W \left( M^{-1}z_1- z_0  + M^{-1}z^\prime \right ),
\end{equation}
which proves that $p(z_1 | z_0) $  depends on $z_1$ and  $z_0$ just on the form $M^{-1}z_1- z_0 $. For definiteness let us define the function 
\begin{equation}
\label{W1}
W_1 (z) = \int d z^\prime W(z^\prime ) W \left (z  + M^{-1}z^\prime \right )
\end{equation}
so that
\begin{equation}
p(z_1 | z_0) = W_1 \left ( M^{-1}z_1- z_0  \right ) .
\end{equation}
 
 \bigskip

To proceed via induction now  we assume that after $j$ measurements $p_j (z_j | z_0)$ is of the form 
\begin{equation}
p_j (z_j | z_0) = W_j (M^{-j}z_j - z_0) ,
\end{equation}
and we have to demonstrate that $p_{j+1} (z_{j+1} | z_0)$ fulfills the theorem. We begin with
\begin{equation}
p_{j+1} (z_{j+1} | z_0) = \int d z_j  p(z_{j+1} | z_j) p_j (z_j| z_0) ,
\end{equation}
so that 
\begin{equation}
p_{j+1} (z_{j+1} | z_0) = \int d z_j W_1 (M^{-1}z_{j+1} - z_j ) W_j (M^{-j}z_j - z_0 ) ,
\end{equation}
that after the unit-Jacobian change of variables 
\begin{equation}
z^\prime = M^{-1} z_{j+1} - z_j,  \qquad z_j = M^{-1} z_{j+1} -  z^\prime,
\end{equation}
becomes
\begin{eqnarray}
p_{j+1} (z_{j+1} | z_0) & = & \int d z^\prime W_1 (z^\prime ) \nonumber \\
& \times & W_j \left [ M^{-j} \left (M^{-1}z_{j+1}  - z^\prime \right ) - z_0 \right ]  , \nonumber \\
 & & 
\end{eqnarray}
which clearly shows that $p_{j+1} (z_{j+1} | z_0) $  depends on $z_{j+1}$ and  $z_0$ just on the form $M^{-(j+1)}z_{j+1}- z_0 $
\begin{equation}
p_{j+1} (z_{j+1} | z_0) = W_{j+1} \left ( M^{-(j+1)}z_{j+1} - z_0 \right )  .
\end{equation}
This satisfies the theorem simply defining 
\begin{equation}
\label{WN}
 W_{j+1} (z) =  \int d z^\prime W_1 (z^\prime ) W_j \left (z -M^{-j} z^\prime \right ) .
 \end{equation}
This completes the proof. 

\bigskip

We finally take into account that when the Hamiltonian is time independent, the composition of $N$ consecutive unperturbed evolutions of duration $\tau$ equals a single evolution of time $N \tau$. This is, if $z_N = M (\tau) z_{N-1}$,  $z_{N-1}= M (\tau) z_{N-2}$, $\dots$, $z_1 = M (\tau) z_0$, then $z_N = M^N (\tau) z_0$, and this must be naturally equal to $z_N = M (N \tau) z_0$. Actually, in our case it can be easily checked the identity
\begin{equation}
M^N (\tau ) = M (N \tau ) ,
\end{equation} 
since $ M (\tau)$ is just a rotation of angle $\Omega \tau$ in phase space.

\subsection{Gaussian states}

Let us compute explicitly the function $W_N  (z)$ in the case of squeezed coherent states, where we can take advantage of the fact that the Wigner function of $|\psi \rangle$ is a Gaussian, which by hypothesis is centered at the origin, say 
\begin{equation}
W (z ) = \frac{1}{2\pi \sqrt{ \det C}}  e^{- z^T C^{-1} z/2} , \quad  C =  \frac{1}{2}  \left ( \begin{array}{cc} e^{2r} & 0 \\ 0 & e^{-2r} \end{array} \right ) ,
\end{equation}
where  $r$ is the compression parameter. After the convolutions in Eqs. (\ref{W1}) and (\ref{WN}) it is clear that all $W_j (z)$ are Gaussians centered at the origin, say 
\begin{equation}
\label{WNG}
W_j (z ) = \frac{1}{2\pi \sqrt{ \det C_j}}  e^{- z^T C_j^{-1} z/2} ,
\end{equation}
where $C_j$ is the corresponding covariance matrix 
\begin{equation}
C_j  = \int dz \; z z^T W_j (z ) .
\end{equation}

\bigskip

Let us derive a recursive relation for $C_j$. Starting from Eq. (\ref{WN}), and after a tricky change of variables of the form $z^{\prime \prime} = z - M^{-j}  z^\prime$ we get in few steps to  
\begin{equation}
C_{j+1}  =  C_j + M^{-j} C_1 \left ( M^{-j} \right )^T ,
\end{equation}
that leads to 
\begin{equation}
\label{CN2}
C_N  = \sum_{j=0}^{N-1} M^{-j} C_1 \left ( M^{-j} \right )^T.
\end{equation}

\end{document}